\documentclass[aps,prb,showpacs,amsmath,amssymb]{revtex4}
\usepackage{graphicx}
\usepackage{amsmath}
\usepackage{amssymb}

\def\prl{{Phys. Rev. Lett.\ }}
\def\prb{{Phys. Rev. B\ }}
\def\npb{{Nucl. Phys. B\ }}
\def\ssc{{Solid State Commun.\ }}

\begin{document}
\title{Collective edge modes in fractional quantum Hall systems}
\author{Hoang K. Nguyen$^{1}$, Yogesh N. Joglekar$^{2}$, and 
Ganpathy Murthy$^{1}$}
\affiliation{$^{1}$Department of Physics and Astronomy, University of 
Kentucky, Lexington, Kentucky 40506 \\
$^{2}$ Theoretical Division, Los Alamos National Laboratory, Los Alamos, 
New Mexico 87545}
\date{\today}

\begin{abstract}
Over the past few years one of us (Murthy) in collaboration with R. Shankar 
has developed an extended Hamiltonian formalism capable of describing the 
ground state and low energy excitations in the fractional quantum Hall 
regime. The Hamiltonian, expressed in terms of Composite Fermion operators, 
incorporates all the nonperturbative features of the fractional Hall regime, 
so that conventional many-body approximations such as Hartree-Fock and 
time-dependent Hartree-Fock are applicable. We apply this formalism to 
develop a {\it microscopic} theory of the collective edge modes in 
fractional quantum Hall regime. We present the results for edge mode 
dispersions at principal filling factors $\nu=1/3,1/5$ and $\nu=2/5$ for 
systems with unreconstructed edges. The primary advantage of the method is 
that one works in the thermodynamic limit right from the beginning, thus 
avoiding the finite-size effects which ultimately limit exact diagonalization 
studies.
\end{abstract}
\maketitle

\section{Introduction}
\label{sec:intro}
The bulk fractional quantum Hall effect is by now well-understood as a 
consequence of an interaction-driven incompressible 
state\cite{laugh,jain,fqhereviews} of a two-dimensional electron gas (2DEG) 
in a strong perpendicular magnetic field. It was also realized long 
ago\cite{halperin} that the edges of the bulk system play a fundamental role 
in transport at low frequencies, since the bulk is incompressible and the 
gapless excitations are available only at the edges.  An effective 
description of edge excitations for incompressible fractions was first 
proposed by Wen.\cite{wen} He developed a hydrodynamic theory of a sharp 
edge, which is a realization of the one-dimensional Chiral Luttinger liquid.

Recently, several high-precision experiments have probed the low-frequency 
dynamics of the edge in the fractional quantum Hall (FQH) regime over a range 
of filling factors.\cite{expts,lloyd} Some features of the results, 
such as the observation of a power law in the $I$-$V$ curve which depends on 
the filling factor, and the magnitude of the power-law exponent,\cite{expts} 
are unexpected from the point of view of hydrodynamic theories of the 
edge.\cite{wen,kane} Many explanations have been proposed for these
discrepancies,\cite{dhlee,ana,ulrich,levitov,goldman,wan,mandal,kun} 
including the effects of long-range interactions and edge reconstructions. 
Motivated by these experiments, Wan {\it et al.}\cite{wan} have recently 
explored the effects of edge reconstructions on the behavior of the 
collective modes using numerical exact diagonalization. While important 
qualitatively new results\cite{kun} have been obtained by this research, 
these studies do suffer from finite-size effects.

We want to investigate the physics of FQH edges using the extended 
Hamiltonian theory developed by Shankar and Murthy.\cite{sm1} This approach 
has the advantages that it starts with a microscopic Hamiltonian, permits 
analytic calculations in the thermodynamic limit and also retains the lowest 
Landau level limit. It thus provides a complementary approach to that of 
exact diagonalization studies. In a previous paper\cite{yogesh} we applied 
the extended Hamiltonian theory to study the structure of FQH edges within 
the Hartree-Fock (HF) approximation. In this paper, we further explore the 
edge states beyond mean-field level. We use the time-dependent Hartree-Fock 
(TDHF) approximation which systematically incorporates particle-hole states 
because it is a conserving approximation,\cite{conserving} {\it i.e.} it 
respects the symmetries of the problem.

The structure of our paper is as follows: In the next section, we review the 
essentials of the extended Hamiltonian theory and mean-field theory of FQH 
edges. The time-dependent Hartree-Fock approximation is described in 
Section~\ref{sec:TDHF}. Our numerical results for $\nu=1/3,1/5$ and 
$\nu=2/5$ are presented in Section~\ref{sec:results}. The results for 
$\nu=2/5$ are, to the best of our knowledge, the first in the literature. 
The last section contains conclusions and caveats. Throughout this paper we 
restrict ourselves to unreconstructed edges due to the computational 
limitations of our own.

\section{Extended Hamiltonian theory and Hartree-Fock approximation}
\label{sec:eht}
For a two-dimensional electron gas in a strong perpendicular magnetic field, 
the kinetic energy of electrons is quenched and the single-particle states 
are organized into Landau levels separated by the cyclotron energy. For 
electrons in the lowest Landau level, neglecting the zero point energy, the 
interaction Hamiltonian is given by
\begin{equation}
H_e=\frac{1}{2A}\sum_{\vec{q}}V_{ee}(\vec{q}):\rho_e(\vec{q})\rho_e(-\vec{q}):
\label{eq: H_e}
\end{equation}
where $V_{ee}(\vec{q})$ is the repulsive electron-electron interaction. This 
interaction energy scale is typically several tens of Kelvins, of the same 
order as the cyclotron energy (about 100K). Nevertheless, in the 
{\it integer} quantum Hall regime, the interaction is usually treated as a 
perturbation on the ground-state Slater-determinant of $\nu$ Landau levels 
completely filled. For a {\it fractional} filling factor, this ground state 
is degenerate and the perturbative approach fails.\cite{laugh} The key insight 
of Jain\cite{jain} was that if one thinks in terms of Composite Fermions 
(CFs), which are envisioned as electrons attached to an even number $2s$ of 
flux quanta, one can use the independent particle picture for the Composite 
Fermions. In this approach, a system at filling factor $\nu=p/(2ps+1)$ has 
CFs which see an effective field $B^*=B/(2ps+1)$, which is just right for 
them to fill $p$ CF Landau levels (CF-LLs). Based on this insight, Jain 
constructed excellent wavefunctions for ground state and low-lying excited 
states for the principal filling factors. However in order to calculate 
response functions or finite-temperature properties, one needs a dynamical 
theory. Initial attempts focused on bosonic\cite{bosonic} or 
fermionic\cite{fermionic} variants of the Chern-Simons theory\cite{cs} 
culminating\cite{kalmeyer} in a Chern-Simons theory for $\nu=1/2$, which is a 
Fermi-liquid-like state.\cite{nu-half} These theories, while satisfactory in 
many respects, still did not incorporate some of the nonperturbative aspects 
of the CF, such as the facts that in the lowest-Landau-level limit the 
effective mass of the CF is determined by the interactions alone or that its 
charge is fractional.

To handle these problems, Shankar and Murthy developed the extended 
Hamiltonian theory,\cite{sm1} a detailed account of which can be found in a 
recent review.\cite{rmpus} A key ingredient of this theory is the 
introduction of a new set of coordinates for {\em pseudovortices} in 
addition to  those of electrons. For a FQH liquid at filling factor 
$\nu=p/(2ps+1)$ where $p$ and $s$ are integers, each electron couples with 
$2s$ pseudovortices to form a composite fermion (CF) in the following way
\begin{eqnarray}
\label{eq:R_e}
\vec{R_e}& = &\vec{r}-\frac{l^2}{1+c}\hat{z}\times\vec{\Pi}, \\
\vec{R_v}& = &\vec{r}+\frac{l^2}{c(1+c)}\hat{z}\times\vec{\Pi},
\label{eq:R_v}
\end{eqnarray}
where $c^2=2\nu s$, $l=\sqrt{h/eB}$ and $\vec{r}$ and $\vec{\Pi}$ are the 
position and velocity operators of the composite fermion. The electron 
guiding center $\vec{R_e}$ and the pseudovortex guiding center $\vec{R_v}$ 
satisfy the algebra
\begin{eqnarray}
\label{eq: commutation}
[R_{e\alpha},R_{v\beta}] &= & 0, \\ \nonumber
[R_{e\alpha},R_{e\beta}] &= & -il^2\epsilon_{\alpha\beta}, \\ \nonumber
[R_{v\alpha},R_{v\beta}] & = & +i\frac{l^2}{c^2}\epsilon_{\alpha\beta}.
\end{eqnarray}
Thus, the electron guiding-center coordinates satisfy the magnetic algebra 
with charge $-e<0$, whereas the pseudovortex guiding-center algebra represents 
an object with charge $+ec^2$. The CFs thus have a magnetic algebra charge of 
$e^*=-e(1-c^2)$, showing that they are subject to a reduced magnetic field 
$B^*=B/(2ps+1)$, just right to fill the first $p$ CF-Landau levels, exactly 
as in Jain's picture.\cite{jain} The Hartree-Fock state of CFs with the first 
$p$ Landau levels filled provides a {\em nondegenerate} starting point for 
analytical calculations.

To calculate the matrix elements of these operators in the CF basis, we use 
the single-particle states of the CFs in the reduced effective field $B^{*}$. 
In the Landau gauge, a single-particle state $|nX\rangle$ is characterized by 
the CF-Landau level index $n$ and the guiding-center coordinate $X$. In the
real-space representation, the (unnormalized) single-particle wavefunction is
\begin{equation}
\langle\vec{r}|nX\rangle=e^{iXy/l^{*^2}}e^{-(x-X)^2/2l^{*^2}}H_n
\left[(x-X)/l^{*}\right],
\label{eq: basis}
\end{equation}
where $H_n(x)$ are the Hermite polynomials and $l^{*}=\sqrt{h/eB^{*}}$ is the 
magnetic length in the reduced field $B^{*}$ seen by the CFs. In the 
following, we use $l^{*}=1$ as the unit of length. Using this 
basis, it is straightforward to express the electron density $\rho_e$ and the 
pseudovortex density $\rho_v$ operators in second-quantized notation
\begin{eqnarray}
\label{eq: rho_e}
\rho_e(\vec{q})& = & \sum_{n_{i}X} e^{-iq_x X} d^\dagger_{n_1 X-q_y/2} 
d_{n_2 X+q_y/2} \rho_{n_1n_2}(\vec q),\\
\label{eq: rho_v}
\rho_v(\vec{q})& = & \sum_{\nu X} e^{-iq_x X} d^\dagger_{n_1 X-q_y/2} 
d_{n_2 X+q_y/2} \chi_{n_1n_2}(\vec q),
\end{eqnarray}
where $d_{nX} (d^\dagger_{nX})$ destroys (creates) a CF in the state 
$|nX\rangle$, and $\rho_{n_1n_2}$ ($\chi_{n_1n_2}$) are the plane-wave matrix 
elements for electron guiding center $\vec{R}_{e}$ (pseudovortex guiding 
center $\vec{R}_{v}$).\cite{rmpus} We have now expressed the electron 
density operators and hence the microscopic Hamiltonian for the fractional 
quantum Hall systems in terms of CF operators, $d_{nX}$ and $d^\dagger_{nX}$. 
Therefore the original problem of interacting electrons in the 
{\it fractional} quantum Hall regime has been readily transformed into a 
problem of CFs in the {\it integer} quantum Hall regime where various 
many-body approximations are applicable. But, we must remember that there is 
no free lunch; the price to pay is that the transformed problem is subject to 
special constraints as a result of the introduction of pseudovortex 
coordinates (section~\ref{sec:TDHF}).

We next consider the positive neutralizing background charge $\rho_b(x)$ which 
produces a confining potential near the edge at $x=0$. The corresponding 
Hamiltonian is  
\begin{eqnarray}
\label{eq: H_b}
H_b & = &\sum_{\vec{q}}{V}_{eb}(-\vec{q})\rho_e(\vec{q}) 
\end{eqnarray}
where 
\begin{equation}
\label{eq: V_eb}
V_{eb}(x) = \int^{\infty}_{-\infty} dx'\, V_{eb}(x-x')\rho_b(x'),
\end{equation}
and $V_{eb}(x)$ is the attractive electron-background interaction. We choose 
a background charge density which vanishes linearly over width $W$ near the 
edge: $\rho_b(x)=0$ for $x<-W/2$, 
$\rho_b(x)=-\frac{\rho_0}{2}\left[1+\frac{2x}{W}\right]$ for $|x|<W/2$, and 
$\rho_b(x)=\rho_0$ for $x>W/2$ where $\rho_0$ is the CF density in the 
bulk.\cite{chamon,yogesh} We also assume that the background charge resides 
in the same plane as the 2DEG, $V_{eb}(\vec{q})=-V_{ee}(\vec{q})$.

In the bulk of a FQH liquid at filling factor $\nu=p/(2ps+1)$, $p$ CF Landau 
levels are completely filled. Due to the presence of the confining potential 
close to the edge, different CF-LLs mix together. Since the system is still 
translationally invariant along the $y$-direction, the Landau gauge quantum 
number $X$ is a good quantum number, and only CF-LLs with the same value of 
$X$ mix. We introduce the mixing matrices $\alpha(X)$ as function of $X$
\begin{equation}
d_{nX}= \sum_m \alpha_{n m}(X) a_{mX}
\label{eq: alpha}
\end{equation}
where $a_{mX}$ denote the annihilation operators for the CF quasiparticles 
which result from mixing of different CF-LLs. The total Hamiltonian, 
$H_e+H_b$, represented in terms of new operators $a_{mX}$ and 
$a^\dagger_{mX}$, is decoupled using standard HF approximation~\cite{rmpus} 
and the mean-field solution is chosen to be 
\begin{equation}
\langle a^\dagger_{mX} a_{nX'}\rangle = \delta_{mn} \delta_{XX'} N^F_m(X)
\end{equation}
where $N^F_m(X)$ is the CF-occupation number at guiding center $X$ and 
CF Landau level $m$. Note that with this ansatz for the 
mean-field solution we are explicitly forbidding any structure in the 
$y$-direction. Upon decoupling, the mean-field Hamiltonian becomes
\begin{eqnarray}
H^{HF} & = & \sum_{mn X} H_{mn}(X) a^\dagger_{mX} a_{nX}
\end{eqnarray}
We determine the mixing matrices $\alpha$'s so that the mean-field 
Hamiltonian $H^{HF}$ is diagonal. The process is normally carried out 
iteratively as in our previous work.~\cite{yogesh} The quantities of 
interest in mean-field study are the energy bands $E_m(X)=H_{mm}(X)$, the 
occupation functions $N^F_m(X)$, and the mixing matrices $\alpha(X)$.

\section{Time-dependent Hartree-Fock Approximation}
\label{sec:TDHF}


\subsection{Time-dependent Hartree-Fock Hamiltonian}
\label{subsec:tdhf}
As we mentioned earlier, special care is required in treating the transformed 
Hamiltonian. In particular, the original electronic Hamiltonian knows 
nothing about the pseudovortex density $\rho_v$ and thus is invariant with 
respect to our choice of it. The transformed problem has extra degrees of 
freedom, the pseudovortex coordinates. Exact solutions to this problem will 
automatically be degenerate along the pseudovortex manifold, and will thus 
have a gauge symmetry. However, the mean-field ground state does not have the 
required gauge invariance. The situation is similar in the theory of 
superconductivity, where the Bardeen-Cooper-Schrieffer (BCS) ground state is 
not gauge invariant. Nonetheless gauge-invariant dynamical response functions 
can still be obtained from the mean-field ground state by a so-called 
conserving approximation.\cite{conserving} TDHF approximation turns out to be 
the simplest of the class of conserving approximations, and is standard in 
the context of superconductivity when dealing with the gauge 
symmetry.\cite{conserving} Read was the first to apply the TDHF approximation 
to study the problem of $\nu=1$ bosons,\cite{nu1bosons} which has important 
similarities to the FQH system at filling factor $\nu=1/2$.\cite{nu-half} It 
was subsequently used by Murthy\cite{Ganpathy} to find the collective modes 
for the incompressible fractions using the extended Hamiltonian theory.

In this Section, we will exploit the TDHF approximation to obtain the 
dispersion relations of the edge excitations. The way the conserving nature 
of the TDHF approximation manifests itself is that the spectrum of the TDHF 
Hamiltonian necessarily contains zero eigenvalues, each corresponding to an 
unphysical gauge degree of freedom, while the rest of the spectrum is 
physical.~\cite{Ganpathy} It is important that the unphysical zero-modes are 
identified and excluded and we will discuss the techniques we use for the 
same in the next subsection.

It follows from Eq.(\ref{eq: commutation}) that 
$\left[\rho_e(\vec q),\rho_v({\vec q}')\right]=0.$ The total Hamiltonian, 
$H_e+H_b$, therefore commutes with the pseudovortex density $\rho_v$. Let us 
consider the magnetoexciton operator
\begin{equation}
O_{m_1m_2}(X,q_y)=a^\dagger_{m_1,X-q_y/2}a_{m_2,X+q_y/2}
\label{eq: exciton}
\end{equation}
where $a_{mX}$ is the {\em rotated} annihilation operator related to the 
CF-operator $d$ as in Eq.~(\ref{eq: alpha}). We have explicitly made use of 
the fact that there is translational invariance in the $y$-direction and 
therefore $q_y$ is a good quantum number. Henceforth we shall suppress the 
index $q_y$ and simply write $O_{m_1m_2}(X)$. The equation of motion for the 
magnetoexciton operator is
\begin{equation}
-i \frac{\partial}{\partial t} O_{m_1m_2}(X) = [H,O_{m_1m_2}(X)]
\label{eq: EOM}
\end{equation}
Using the standard HF decoupling to evaluate the right-hand-side we get 
\begin{widetext}
\begin{eqnarray}
\left[H,O_{m_1m_2}(X_0)\right]& \rightarrow&[E_{m_1}(X_0-q_y/2)-
E_{m_2}(X_0+q_y/2)]O_{m_1m_2}(X_0)+[N^F_{m_1,X_0-q_y/2}-N^F_{m_2,X_0+q_y/2}]
\nonumber \\
& &\times\sum_{X\nu k_i k'_i}
\left[h_{k_1k_2'k_1'k_2}(X_0-X,-q_y)-h_{k_1k_2k_1'k_2'}(X-X_0,q_y)\right]
\nonumber \\
& &\times\ \ \alpha^\dagger_{k_1n_1}(X-q_y/2)\alpha_{k_2n_2}(X+q_y/2)
\alpha^\dagger_{k_1'm_2}(X_0+q_y/2)\alpha_{k_2'm_1}(X_0-q_y/2)O_{n_1n_2}(X)
\label{eq: HO}
\end{eqnarray}
where
\begin{equation}
h_{k_1k_2k_1'k_2'}(X,q_y)=\int\frac{dq_x}{2\pi}V_{ee}(\vec q)\rho_{k_1k_2}
(\vec q)\rho_{k_1'k_2'}(-\vec q)e^{-iq_xX}
\label{eq: h}
\end{equation}
and $\mu=(m_1,m_2)$, $\nu=(n_1,n_2)$ {\it etc.} We define a vector 
$\Psi_\mu(X)$ corresponding to the operator
\begin{equation}
O_\Psi = \sum_{\mu X} \Psi_\mu(X) O_\mu(X).
\label{eq: O}
\end{equation}
In the space spanned by $O_\Psi$, the TDHF Hamiltonian reads
\begin{eqnarray}
{\cal H}(\mu,\nu;X_0,X;q_y) &=&\delta_{m_1n_1}\delta_{m_2n_2} 
[E_{m_1}(X_0-q_y/2)-E_{m_2}(X_0+q_y/2)]+
[N^F_{m_1,X_0-q_y/2}-N^F_{m_2,X_0+q_y/2}]\nonumber \\
& &\times\sum_{k_i,k'_i}
\left[h_{k_1k_2'k_1'k_2}(X_0-X,-q_y)-h_{k_1k_2k_1'k_2'}(X-X_0,q_y)\right]
\nonumber \\
& &\times\ \ \alpha^\dagger_{k_1n_1}(X-q_y/2)\alpha_{k_2n_2}(X+q_y/2)
\alpha^\dagger_{k_1'm_2}(X_0+q_y/2)\alpha_{k_2'm_1}(X_0-q_y/2)
\label{eq: TDHFHam}
\end{eqnarray}
\end{widetext}

The TDHF Hamiltonian is the key quantity containing information about the 
collective modes in the system. Upon diagonalization, we obtain the energy 
spectrum of excitations, and the left ($\psi^{L}$) and right ($\psi^{R}$) 
eigenvectors. Direct inspection of Eqs.(\ref{eq: h}) and (\ref{eq: TDHFHam}) 
shows that $\cal H$ changes sign, after a suitable reshuffling of indices, as 
$q_y$ does. That means the spectrum at a $q_y$ contains information about the 
spectrum at $-q_y$ as well. The left- and right-eigenvectors are identical 
(modulo reshuffled indices) for both $\pm q_y$. We are interested in the 
gapless excitations across the edge. These modes are in the regime of low 
energy and momentum, and our task at hand is to identify these modes from the 
energy spectrum.

\subsection{Identification of the Physical Modes}
\label{subsec:ipm}
Due to the gauge symmetry of the original Hamiltonian the TDHF Hamiltonian 
$\cal H$ necessarily has zero eigenvalues, even for a nonuniform system. In 
the exact solution of the original Hamiltonian each energy level is 
infinitely degenerate since it costs no energy to rotate among 
gauge-connected eigenstates. The zero modes of $\cal H$ are thus unphysical 
and we have to discard them from its spectrum. We stress that the spurious 
modes have precisely zero energy only when the full set of infinite CF-LLs 
are taken into account. Obviously, in computations, though one tries to 
include as many CF-LLs as possible, that number cannot be infinite. 
Therefore, the technical problem of how to separate the unphysical modes from 
the low-lying physical ones arises. There are two different ways to achieve 
this. The first one is to monitor the behavior of the energy of a mode. As 
more CF-LLs are added, the energy of an unphysical mode gets closer to zero 
while that of a physical mode becomes stable. In the bulk case the unphysical 
modes were singled out successfully using this procedure.\cite{Ganpathy} The 
second is to resort to the electron density correlation function. Since the 
electron density operator is gauge-invariant, unphysical modes should 
decouple from its correlation functions as more CF-LLs are added.

In the edge case, the first method is not practical because the energy 
spectrum of $\cal H$ contains every mode at a given $q_y$ including many 
unphysical modes which have high $q_x$ and therefore require a very large 
number of CF-LLs for their detection. We shall instead use the second method. 
We define the electron density correlation function 
\begin{equation}
S(X_0,X'_0;q_y,\omega) = -i \langle \rho_e(X_0,q_y,\omega)
\rho_e(X_0',-q_y,-\omega) \rangle
\label{eq: S}
\end{equation}
as the ground-state expectation of a four-fermion operator. A detailed but 
straightforward calculation gives rise to the following expression for the 
density correlator in terms of known quantities,
\begin{widetext}
\begin{eqnarray}
S(X_0,X'_0;q_y,\omega)  & = & \frac{A}{2\pi} \sum_{\alpha,k_i}\left[
\sum_{\mu X}\rho_{m_1m_2}(X_0-X,q_y)\psi^{R(\alpha)}_{k_1k_2}(X,q_y)
\alpha^\dagger_{m_1k_1}(X-q_y/2)\alpha_{m_2k_2}(X+q_y/2) \right]\nonumber \\
 & & \times\left[\sum_{\nu X'} \rho_{n_1n_2}(X'_0-X',-q_y)
\psi^{L(\alpha)}_{k_4k_3}(X',-q_y)\alpha^\dagger_{n_1k_3}(X'+q_y/2)
\alpha_{n_2k_4}(X'-q_y/2)\right]\nonumber \\
& & \times\left[N^F_{k_3}(X'+q_y/2)-N^F_{k_4}(X'-q_y/2)\right]
\left(\omega-E_\alpha+i\eta\,sign(E_\alpha)\right)^{-1}
\label{eq: correlator}
\end{eqnarray}
\end{widetext}
where $\rho_\mu(X,q_y)$ is the 1D Fourier transform of the density matrix 
\begin{equation}
\rho_\mu(X,q_y)= \int \frac{dq_x}{2\pi}\, e^{i q_xX}\rho_\mu(q_x,q_y)
\end{equation}

As handy as it looks, Eq.~(\ref{eq: correlator}) still has a minor technical 
problem. It describes how strongly a mode couples with the electron density; 
in other words, while it does discriminate between physical and unphysical 
modes, it does not discriminate between the bulk and edge modes. We note that 
physical edge modes at low momentum and energy, the regime of interest, 
should become stable as more and more CF-LLs are added. In a further 
refinement we exclude the high-momentum modes by the following filtering 
procedure. Let us consider, instead of the pure electron density $\rho_e$ 
used in Eq.(\ref{eq: correlator}), a filtered operator concentrated near the 
edge
\begin{equation}
\rho^f_\mu(X_0,q_y)=\sum_X e^{-(X-X_0)^2/2l^2_f} \ \rho_\mu(X,q_y)
\label{eq: rho_f}
\end{equation}
that cuts off all fluctuations across the edge with length-scales longer than 
$l_f$. The choice of Gaussian filtering in Eq.(\ref{eq: rho_f}) is convenient 
for calculations. The results we present in the following section are based 
on the choice $l_f=l^*$, which seems to work well. The filtered density 
correlator, $S^f(X_0,X'_0;q_y,\omega) = -i\langle\rho^f_e(X_0,q_y,\omega)
\rho^f_e(X'_0,-q_y,-\omega) \rangle$ satisfies exactly the same formula 
(\ref{eq: correlator}) except that the electronic density matrix elements 
$\rho_\mu(X,q_y)$ are replaced by the filtered density ones 
$\rho^f_\mu(X,q_y)$, defined as
\begin{equation}
\rho^f_\mu(X,q_y)= \int \frac{dq_x}{2\pi}\,e^{-q_x^2l_f^2/2+iq_x X}
\rho_\mu(q_x,q_y)
\end{equation}

\section{Results}
\label{sec:results}
In the following calculations, we vary the confining potential at the edge 
by changing the width $W$ over which the background charge vanishes linearly 
from its bulk value to zero. We consider a short-ranged Gaussian interaction 
\begin{equation}
V_{ee}(\vec q)= V_0 e^{-q^2\, \lambda^2/2}= -V_{eb}(\vec q)
\label{eq:gaussian}
\end{equation}
with range $\lambda=l^{*}$, and a Thomas-Fermi interaction
\begin{equation}
V_{ee}\vec q) = \frac{2\pi e^2}{\sqrt{q^2+q_{TF}^2}}=-V_{eb}(\vec q)
\label{eq:tf}
\end{equation}
with $q_{TF}l^{*}=0.4$, which becomes truly long ranged as 
$q_{TF}\rightarrow 0$. The Thomas-Fermi case is considerably more difficult 
for the HF part of the calculation. As $q_{TF}$ decreases, it takes more 
iterations for the HF to converge and the presence of the edge makes itself 
felt deeper inside the bulk. We do not find any qualitative difference in the 
edge collective modes between the two cases. This being the case, we 
concentrate on the short-range Gaussian interactions for $\nu=1/5$ and 
$\nu=2/5$.

Deep in the bulk, the quantum Hall liquid is uniform and the structure of 
CF-LLs and excitations in TDHF is well-understood.~\cite{Ganpathy} For 
numerical purposes we focus on the region around the edge, taking $8l^{*}$ in 
the bulk side and $2l^{*}$ in the empty side. The lattice spacing is 
$\Delta x=0.1l^{*}$, resulting in 100 sites. Typically, 8 to 10 CF-LLs are 
considered. The TDHF Hamiltonian size is of the order of a few thousand, 
making the diagonalization feasible. Notice that $\Delta x$ sets the lower 
bound on resolution for $q_y$ (recall that the magnetoexciton contains CF 
creation and destruction operators separated by distance $q_y$.) We believe 
that our choice of $\Delta x=0.1l^{*}$ is appropriate as the natural 
length-scale in the problem is $l^{*}$. Reducing $\Delta x$, while allowing 
us to explore smaller values of $q_y$, would require more computational 
resources.
\begin{figure}[htb]
\begin{center}
\begin{minipage}{20cm}
\begin{minipage}{9cm}
\includegraphics[scale=0.4]{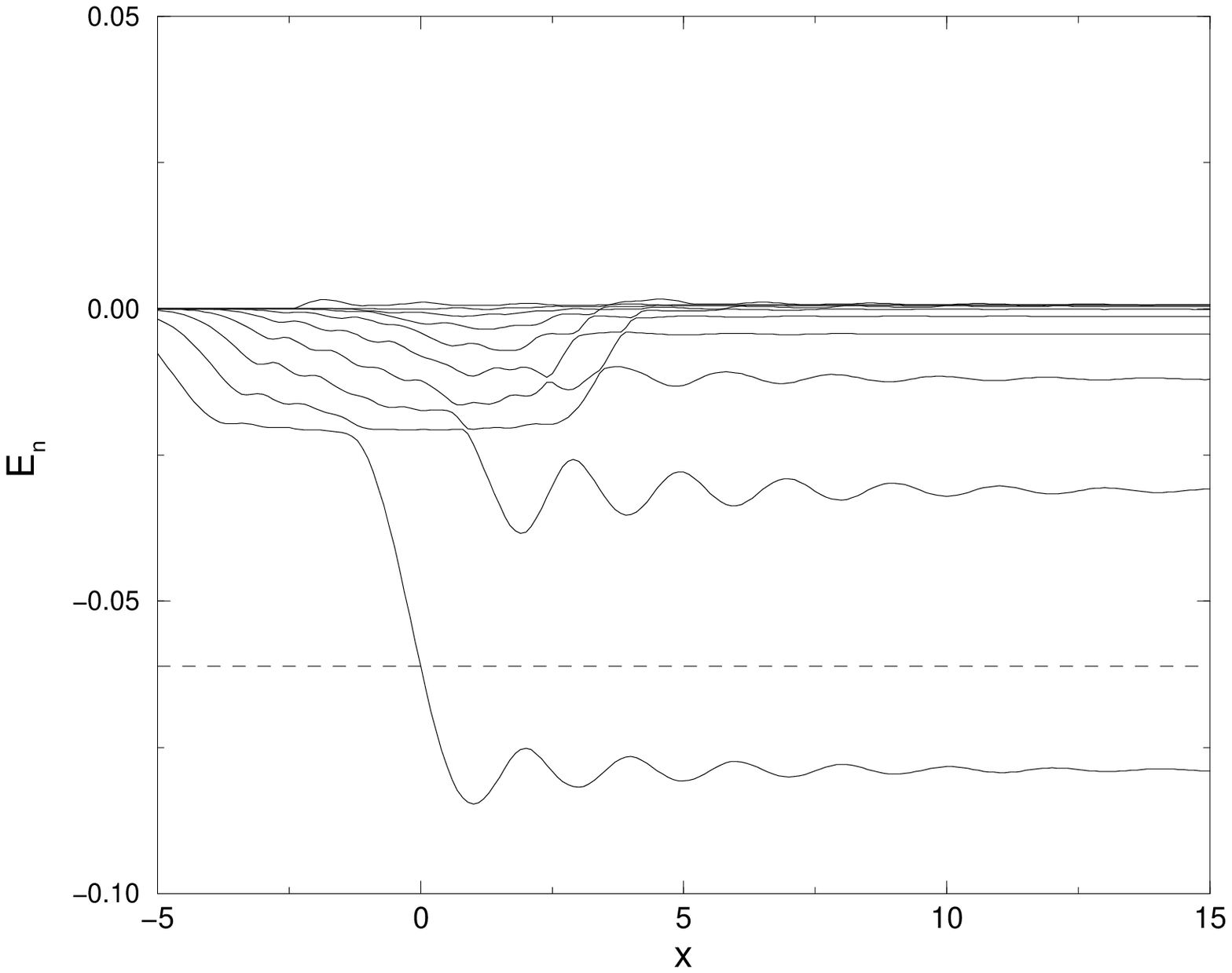}
\end{minipage}
\begin{minipage}{9cm}
\includegraphics[scale=0.4]{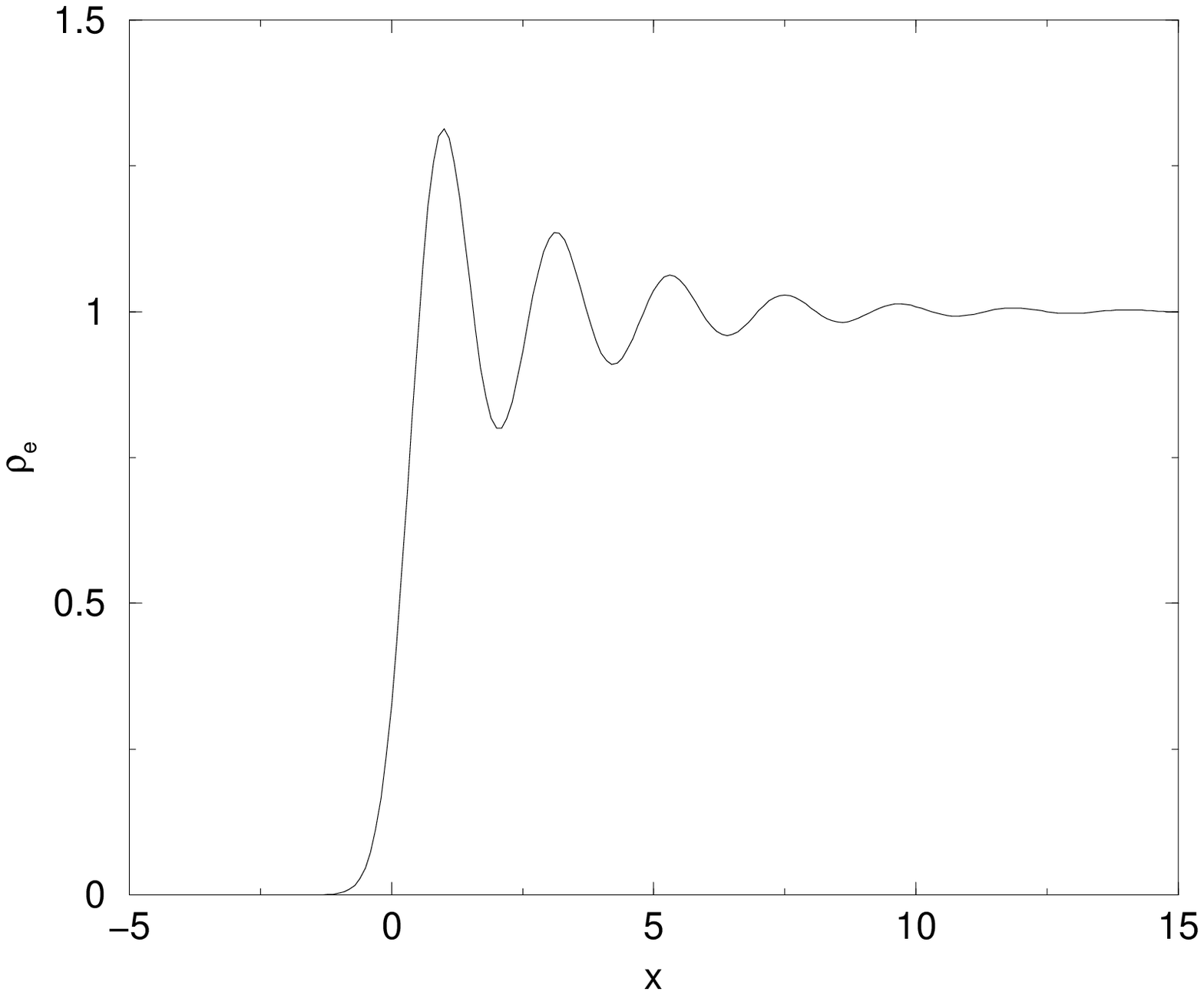}
\end{minipage}
\end{minipage}
\caption{Typical mean-field energy bands (left) and electron density profile 
(right) at the edge for $\nu=1/3$. We have used the Gaussian interaction, 
$W=0$ and 10 CF-LLs. The energy is measured in units of $e^2/\epsilon l^*$ 
and length is measured in units of $l^{*}$ (Note that our interaction is not 
normalized). Only one CF-LL is occupied in the bulk, as signified by the bulk 
electron density $\rho_e=1$ in the right panel; the dotted line in the left 
panel denotes the chemical potential.}
\label{fig: HF.nu13}
\end{center}
\end{figure}

First we present results for the $\nu=1/3$ case, which corresponds to 
$s=1=p$ or $c=\sqrt{2/3}$. In this case, the ground state in the bulk has 
only the lowest CF-LL filled. We solve the HF mean-field equations to obtain 
the energy bands, the mixing matrices, and the CF occupation profile. 
Figure~\ref{fig: HF.nu13} shows typical mean-field energy bands (left) and 
electron density profile (right) near the edge for a Gaussian interaction, 
with $W=0$ and 10 CF-LLs. From the mean-field results the TDHF Hamiltonian 
$\cal H$, Eq.(\ref{eq: TDHFHam}), is created and diagonalized. A direct 
inspection of $\cal H$ reveals that only the excitations from the lowest 
CF-LL to higher ones are relevant to the edge modes. We thus need to keep 
only elements in $\cal H$ that have $m_1=0$ or $m_2=0$. The size of the TDHF 
Hamiltonian with 10 CF-LLs is $100\times 19=1,900$ rows (and columns). While 
diagonalizing a matrix this size is easy, the most time-consuming part of the 
computation is the evaluation of the matrix $\cal H$ due to the quadruple sum 
in Eq.~(\ref{eq: TDHFHam}). Although the TDHF Hamiltonian is not symmetric, 
remarkably, our diagonalization results find that it only possesses 
{\em real} eigenvalues, meaning every mode is {\em long-lived} as normally 
expected. Upon diagonalization, the set of left- and right- eigenvectors are 
used to compute the filtered density correlator. 

\begin{figure}[htb]
\includegraphics[scale=0.4]{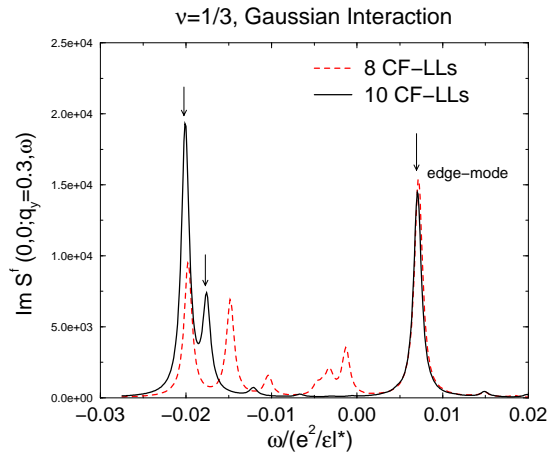}
\caption{Spectral function of the local electron density correlator, 
Im $S^f(0,0;q_y=0.3,\omega)$ for a $\nu=1/3$ system with Gaussian 
interaction and $W=0$. The dotted (solid) lines show results with 8 (10) 
CF-LLs. The arrows indicate poles of physical edge modes. Notice that 
the spurious modes lying between the arrows are strongly suppressed as the 
number of CF-LLs is increased from 8 to 10, whereas the physical edge mode 
with positive energy remains essentially unchanged in both position and 
weight.}
\label{fig: spectral.nu13}
\end{figure}
Figure~\ref{fig: spectral.nu13} shows the imaginary part of the local 
filtered correlator $S^f(0,0;q_y=0.3,\omega)$. It is evident from the picture 
that while there are several peaks in Im $S^f$ at 8 CF-LLs, many of them at 
low energy disappear at 10 CF-LLs. These peaks represent spurious modes which 
decouple from the electron density in the limit of infinite number of CF-LLs. 
The arrows indicate peaks corresponding to the physical modes which are 
{\em stable} as more CF-LLs are added. The peak at positive energy is the 
edge mode; it changes little in position and weight as the number of CF-LLs 
is increased. The gradual disappearance of unphysical modes and the stability 
of the physical modes with respect to increasing number of CF-LLs are strong 
indications that our results, although obtained using a finite number of 
CF-LLs, are reliable. 

\begin{figure}[htb]
\begin{center}
\begin{minipage}{20cm}
\begin{minipage}{9cm}
\includegraphics[scale=0.4]{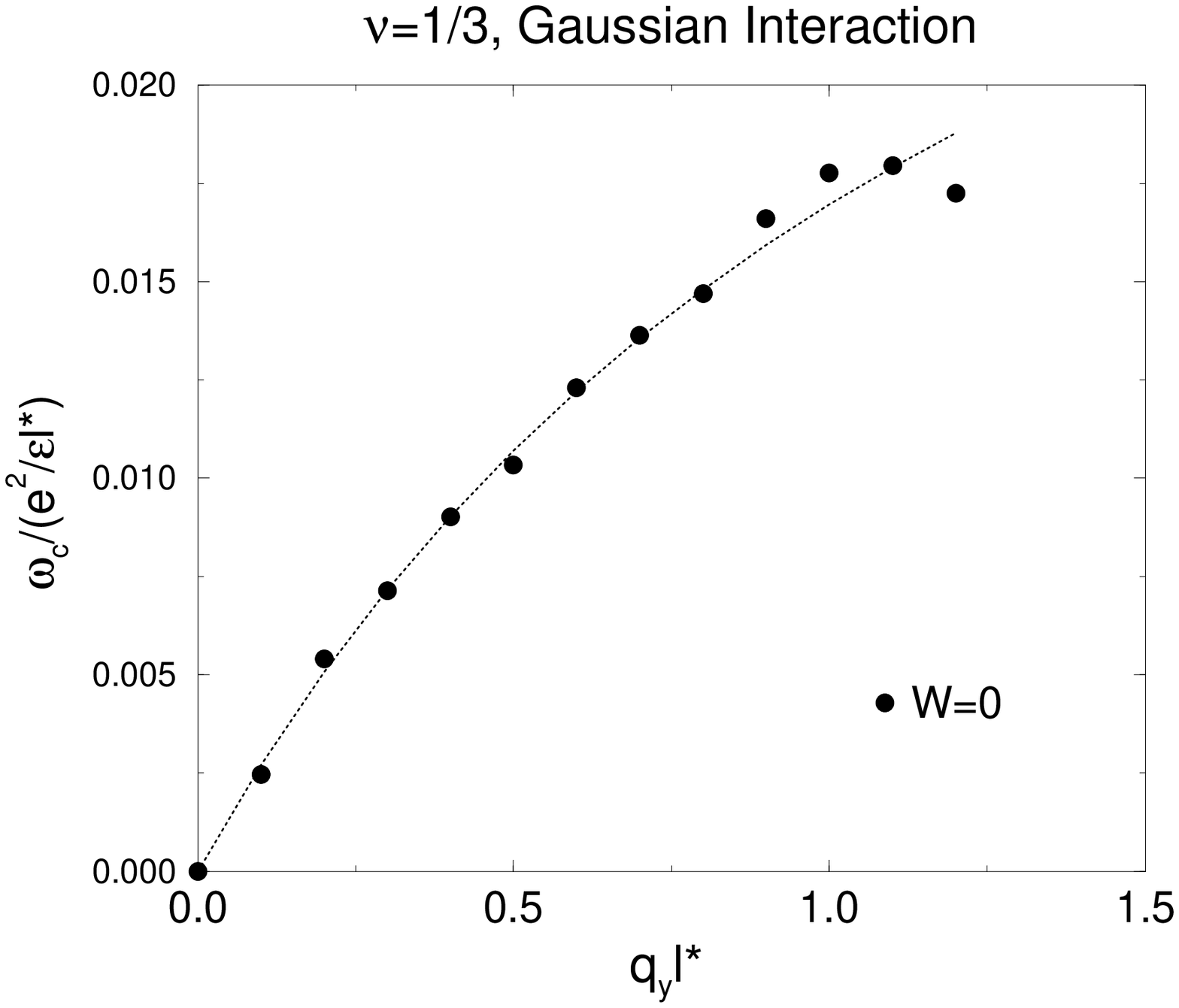}
\end{minipage}
\begin{minipage}{9cm}
\includegraphics[scale=0.4]{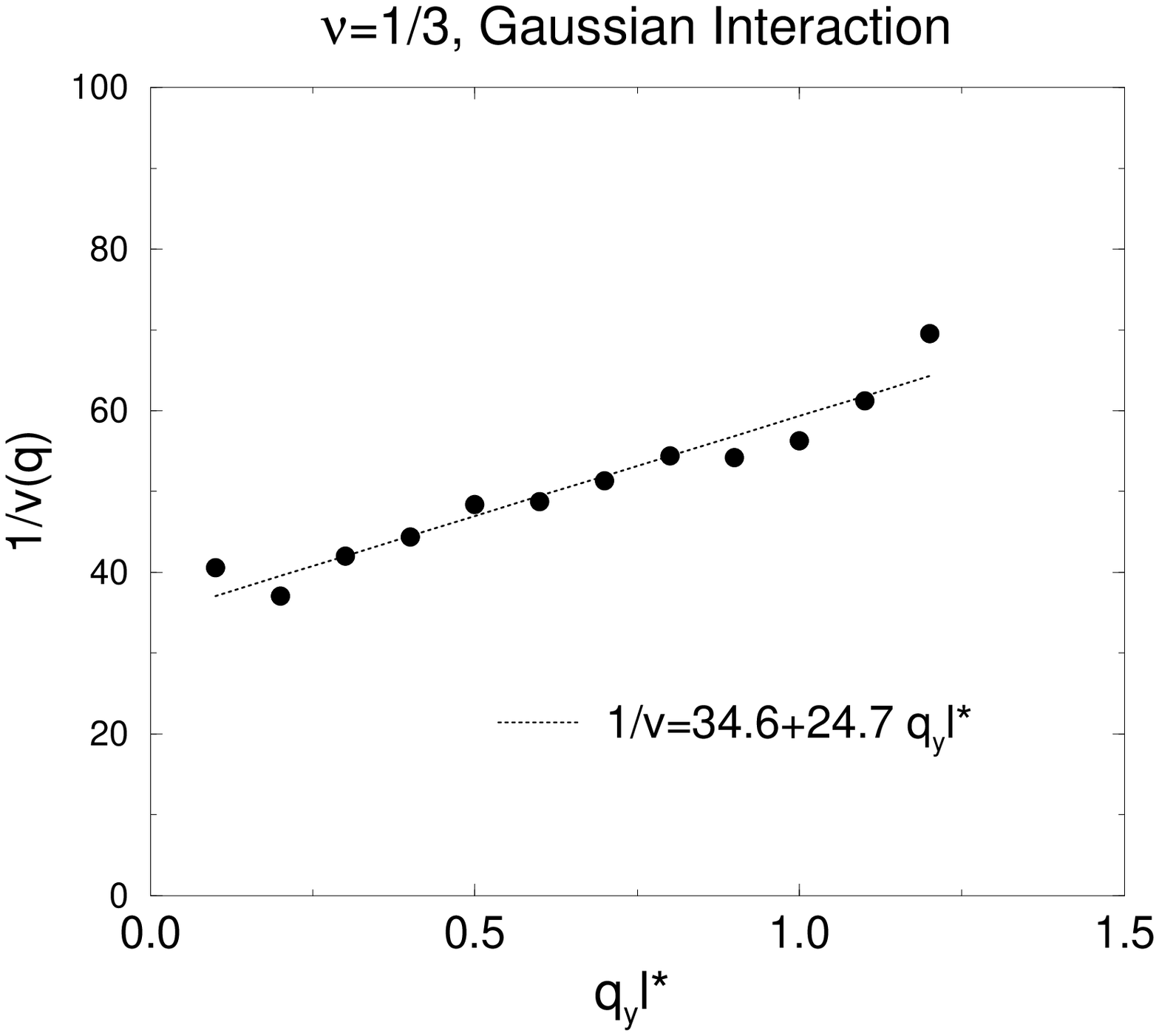}
\end{minipage}
\end{minipage}
\caption{Dispersion (left) and inverse velocity (right) of $\nu=1/3$ edge 
mode for a system with Gaussian interaction, $W=0$, and 10 CF-LLs. In the 
left panel, the dotted line is the naive mean-field result whereas the solid 
line is the dispersion obtained from TDHF approximation. The right panel 
shows the momentum-dependent inverse velocity $1/v(q)$ which fits the 
heuristic form (\ref{eq: inverse_velocity}). The suppression of velocity at 
large $q$ is due to the decreasing electron density near the edge.}
\label{fig: dispersion.nu13}
\end{center}
\end{figure}
By carrying out this procedure for different values of $q_y$ we obtain the 
dispersion relation for the $\nu=1/3$ edge mode, which is shown in 
Fig.~\ref{fig: dispersion.nu13}. The TDHF approximation, which takes into 
account the collective particle-hole pair states, reduces the excitation 
energy below its mean-field value and is shown with a dotted line. As 
expected, the edge mode is gapless, and at small momenta the dispersion is 
linear, $\omega_c(q)= v_0q_y$, where $v_0$ is the cyclotron velocity of the 
CF along the edge. For larger momenta, we assume a momentum-dependent 
velocity $v(q)$. To see heuristically what the natural form of this velocity 
might be, note that as $q_y$ increases we are making excitations deeper into 
the empty region. In this region the electron density is smaller than that in 
the bulk and thus the effective field $B^{*}$ seen by the CFs must be larger. 
This means the velocity of excitations must be smaller. Assuming a linear 
variation of the electron density and thus the effective field near the edge 
(Fig.~\ref{fig: HF.nu13}) we arrive at the following form,
\begin{equation}
\frac{1}{v(q)}=\frac{1}{v_0}+aq_y
\label{eq: inverse_velocity}
\end{equation}
where $v_0=v(q=0)$ and $a$ is a constant. The right panel in 
Fig.~\ref{fig: dispersion.nu13} shows the inverse CF velocity corresponding 
to the dispersion shown in the left panel. Given that there is only one free 
parameter $a$, the data seem to fit our ansatz (\ref{eq: inverse_velocity}) 
very well.

\begin{figure}[htb]
\begin{center}
\begin{minipage}{20cm}
\begin{minipage}{9cm}
\includegraphics[scale=0.4]{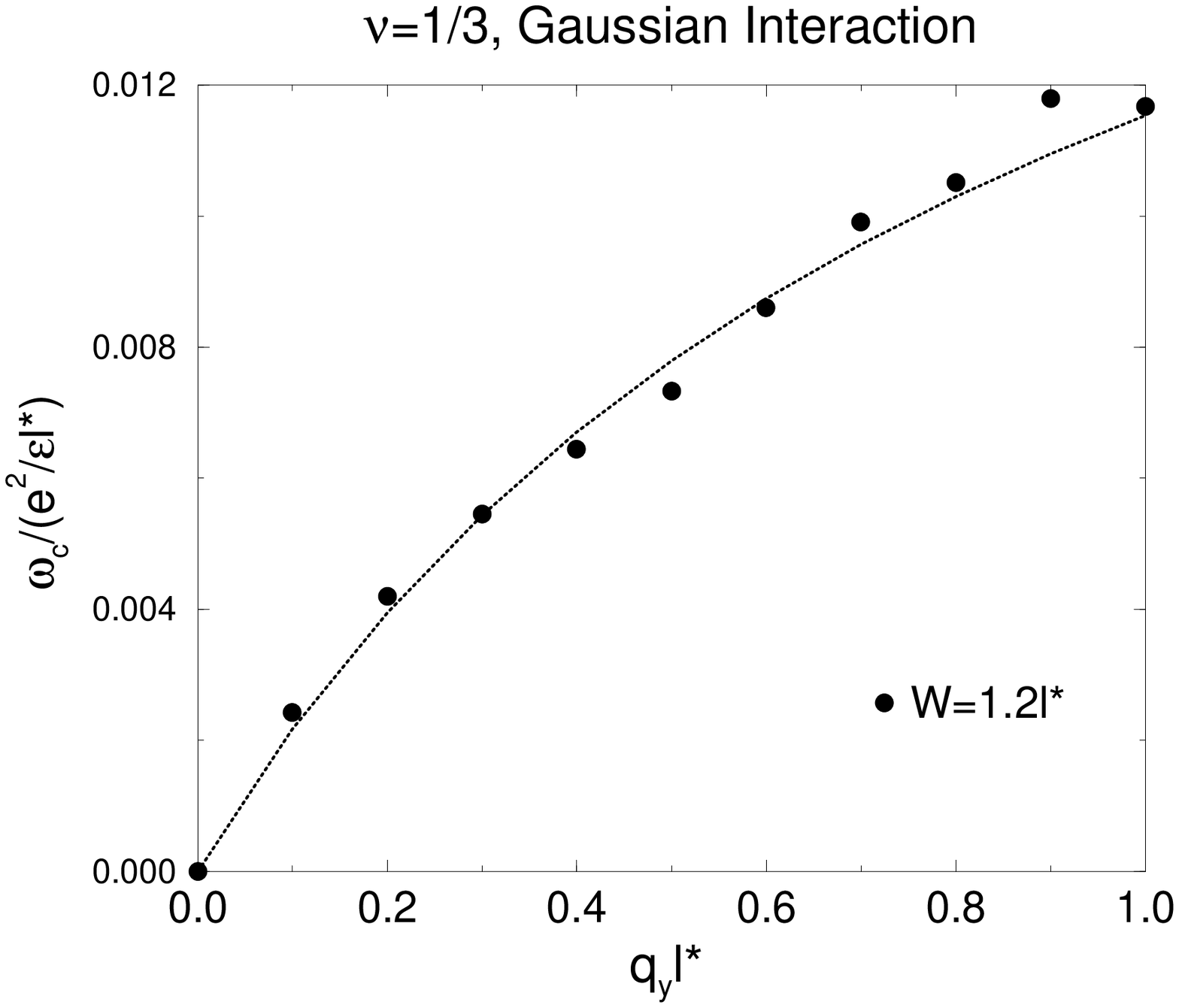}
\end{minipage}
\begin{minipage}{9cm}
\includegraphics[scale=0.4]{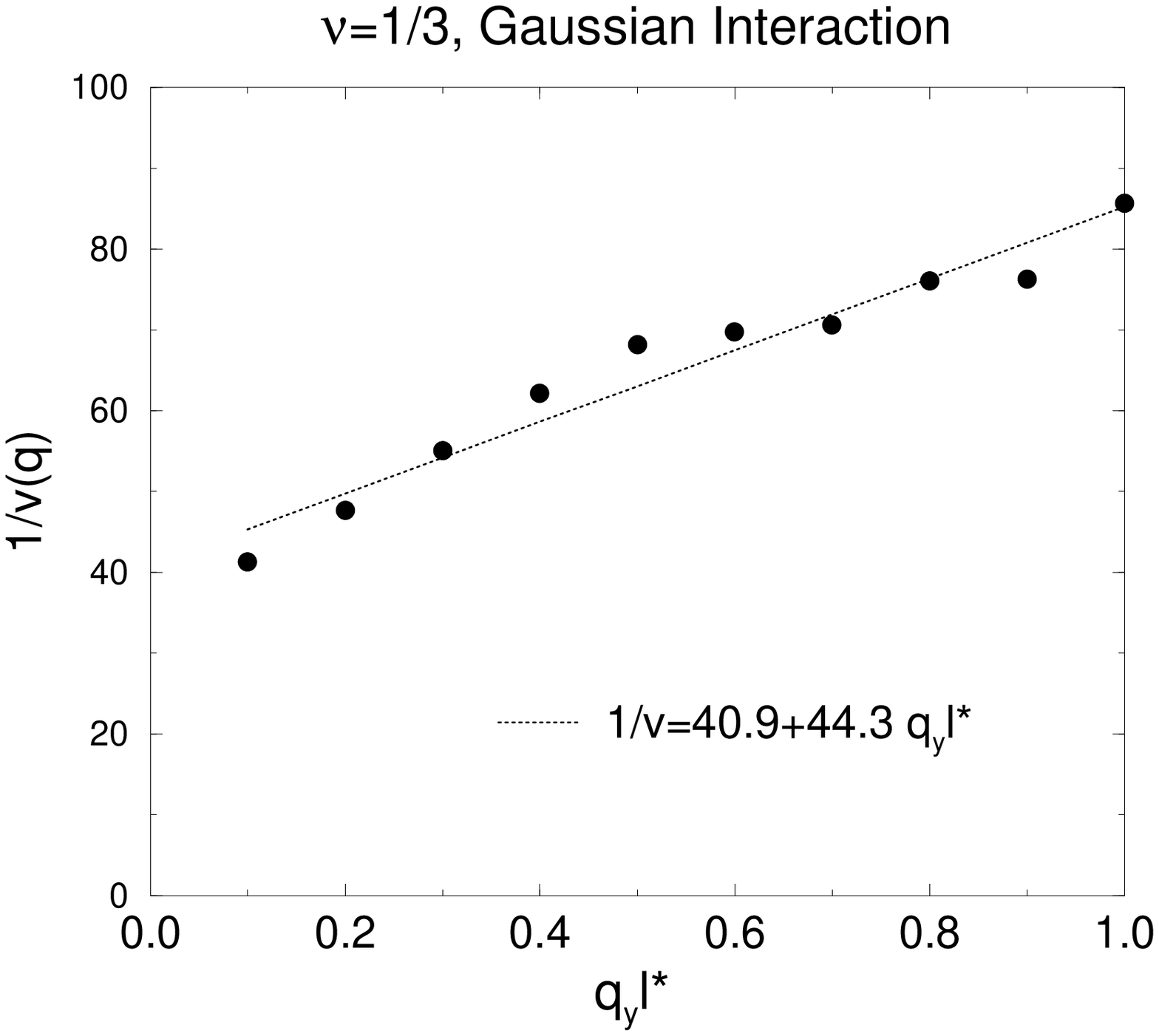}
\end{minipage}
\end{minipage}
\caption{Dispersion (left) and inverse CF velocity (right) of the $\nu=1/3$ 
edge mode with Gaussian interaction, 10 CF-LLs and $W=1.2l^{*}$. Note that 
the edge mode softens as $W$ increases (Fig.~\ref{fig: dispersion.nu13}).}
\label{fig: dispersion.nu13.W12}
\end{center}
\end{figure}

Now we examine how the collective mode dispersion changes as the width of 
the confining potential $W$ is increased. We remind the reader that as $W$ 
increases, the $\nu=1/3$ edge becomes susceptible to reconstruction 
irrespective of the range of electron-electron 
interaction.\cite{chamon,yogesh} Therefore we expect a substantial softening 
of the collective mode dispersion as a precursor to the edge reconstruction. 
Figure~\ref{fig: dispersion.nu13.W12} shows the dispersion (left) and the 
inverse CF velocity (right) for $W=1.2l^{*}$; indeed, as expected, the 
dispersion for $W=1.2l^{*}$ is softer than that for $W=0$ 
(Fig.~\ref{fig: dispersion.nu13}). 

\begin{figure}[htb]
\begin{center}
\begin{minipage}{20cm}
\begin{minipage}{9cm}
\includegraphics[scale=0.4]{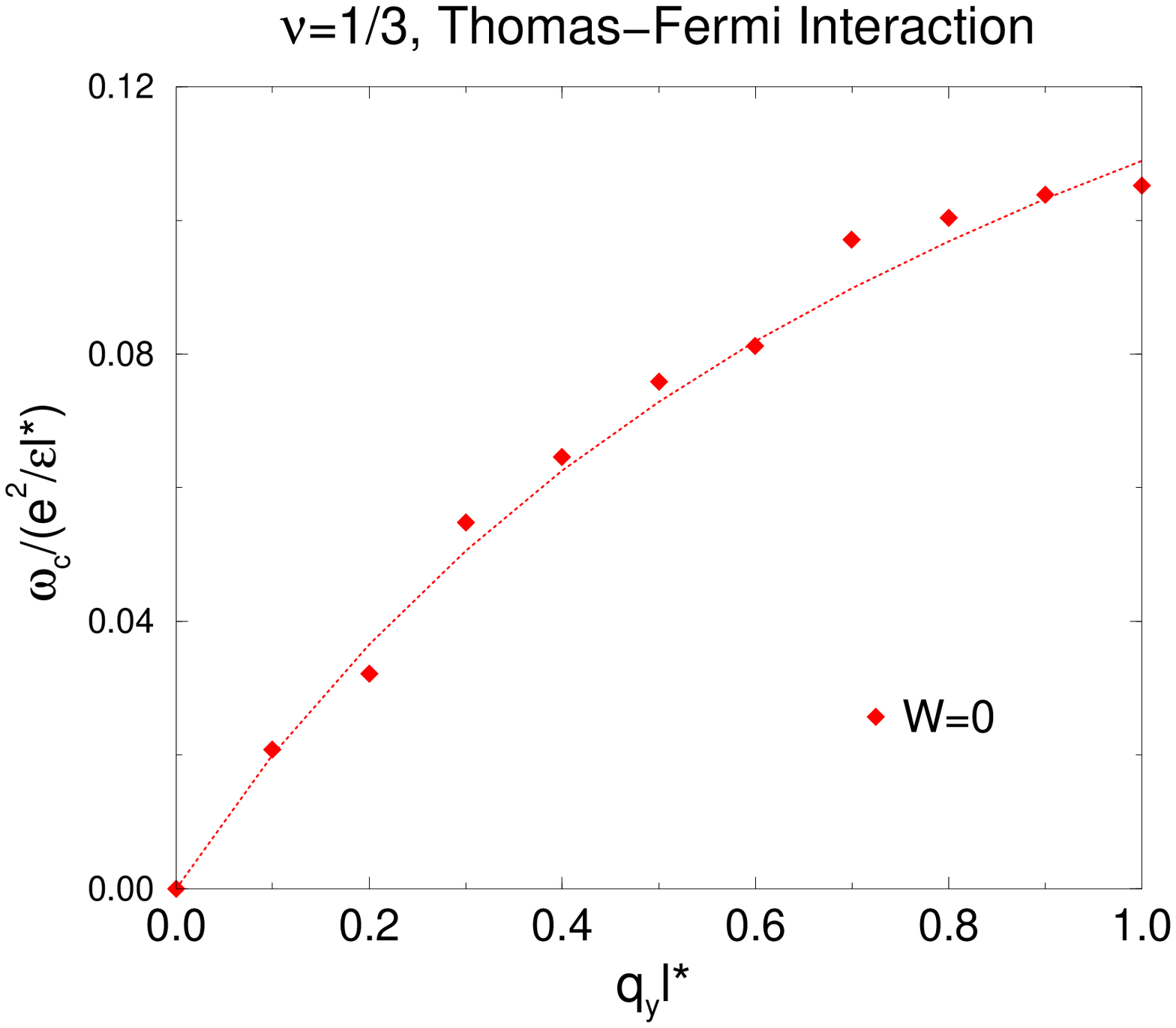}
\end{minipage}
\begin{minipage}{9cm}
\includegraphics[scale=0.4]{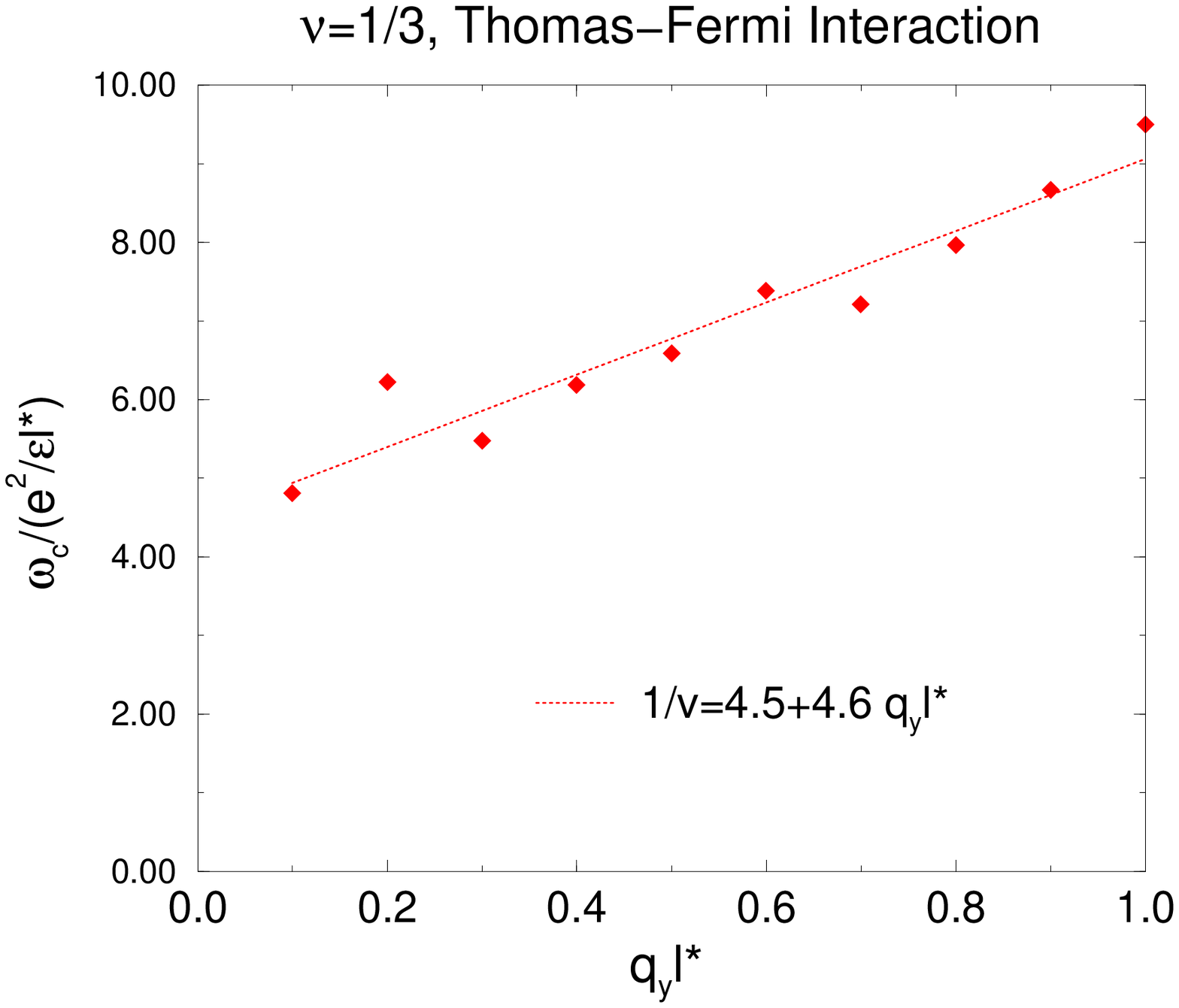}
\end{minipage}
\end{minipage}
\caption{Dispersion (left) and inverse CF velocity (right) for $\nu=1/3$ 
state with Thomas-Fermi interaction, $W=0$, and 10 CF-LLs. Note that the 
dispersion is similar to that in the case of Gaussian interactions and our 
ansatz (\ref{eq: inverse_velocity}) for inverse velocity works for 
long-ranged interactions as well.}
\label{fig: dispersion.TFnu13}
\end{center}
\end{figure}

To illustrate the effect of the range of interaction on the edge modes, we 
also study the $\nu=1/3$ case with Thomas-Fermi interaction. In this case, it 
is harder to filter out the spurious modes than in the case with short-ranged 
Gaussian interactions; however, our scheme still works reasonably well. 
Figure~\ref{fig: dispersion.TFnu13} shows the edge-mode dispersion (left) and 
inverse CF velocity (right) with $W=0$ and 10 CF-LLs. These results are 
qualitatively similar to those obtained using Gaussian interaction. The 
overall change in the edge-mode energy-scale is due to our use of 
un-normalized Gaussian interaction and has no fundamental significance.

\begin{figure}[htb]
\begin{center}
\begin{minipage}{20cm}
\begin{minipage}{9cm}
\includegraphics[scale=0.4]{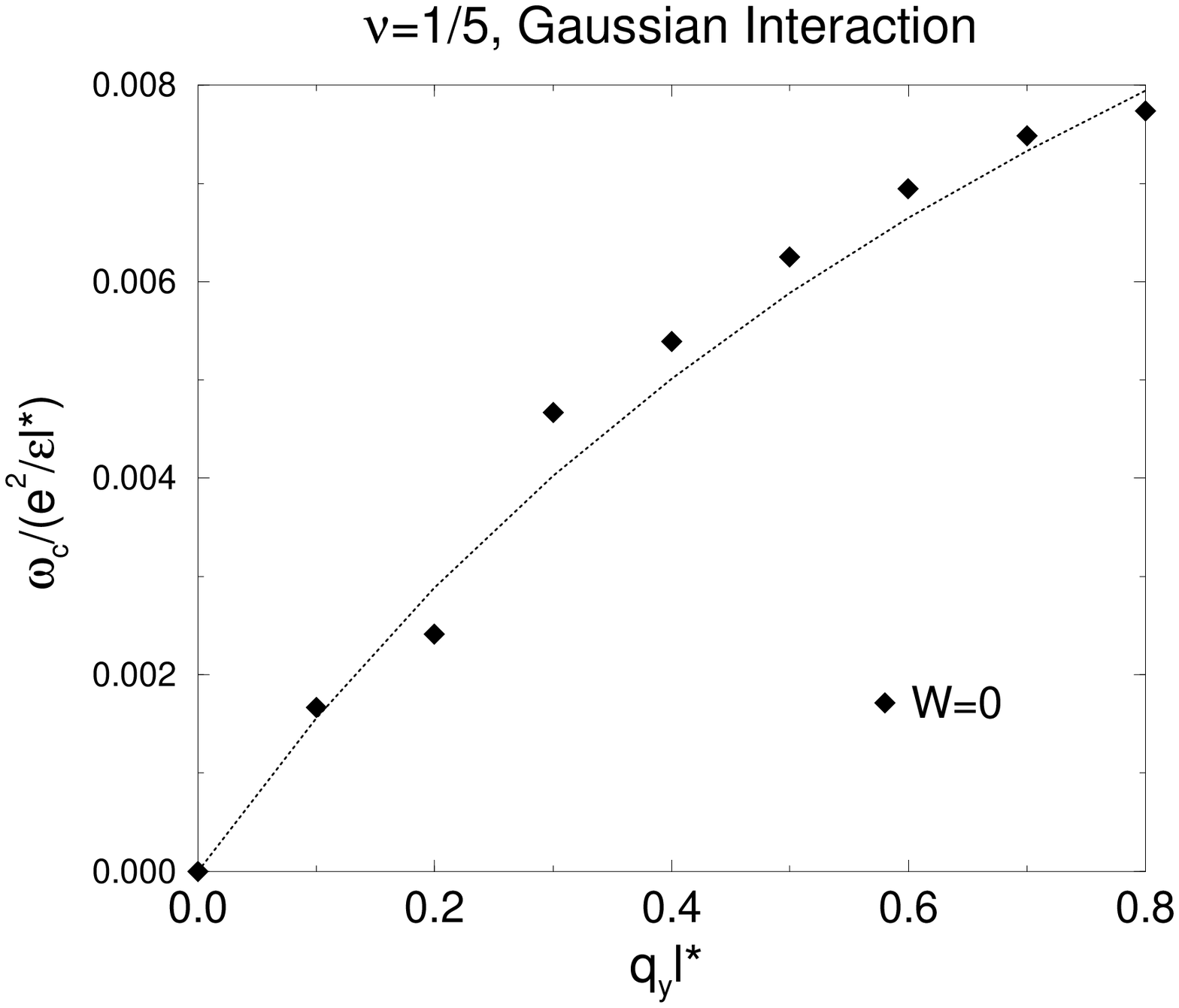}
\end{minipage}
\begin{minipage}{9cm}
\includegraphics[scale=0.4]{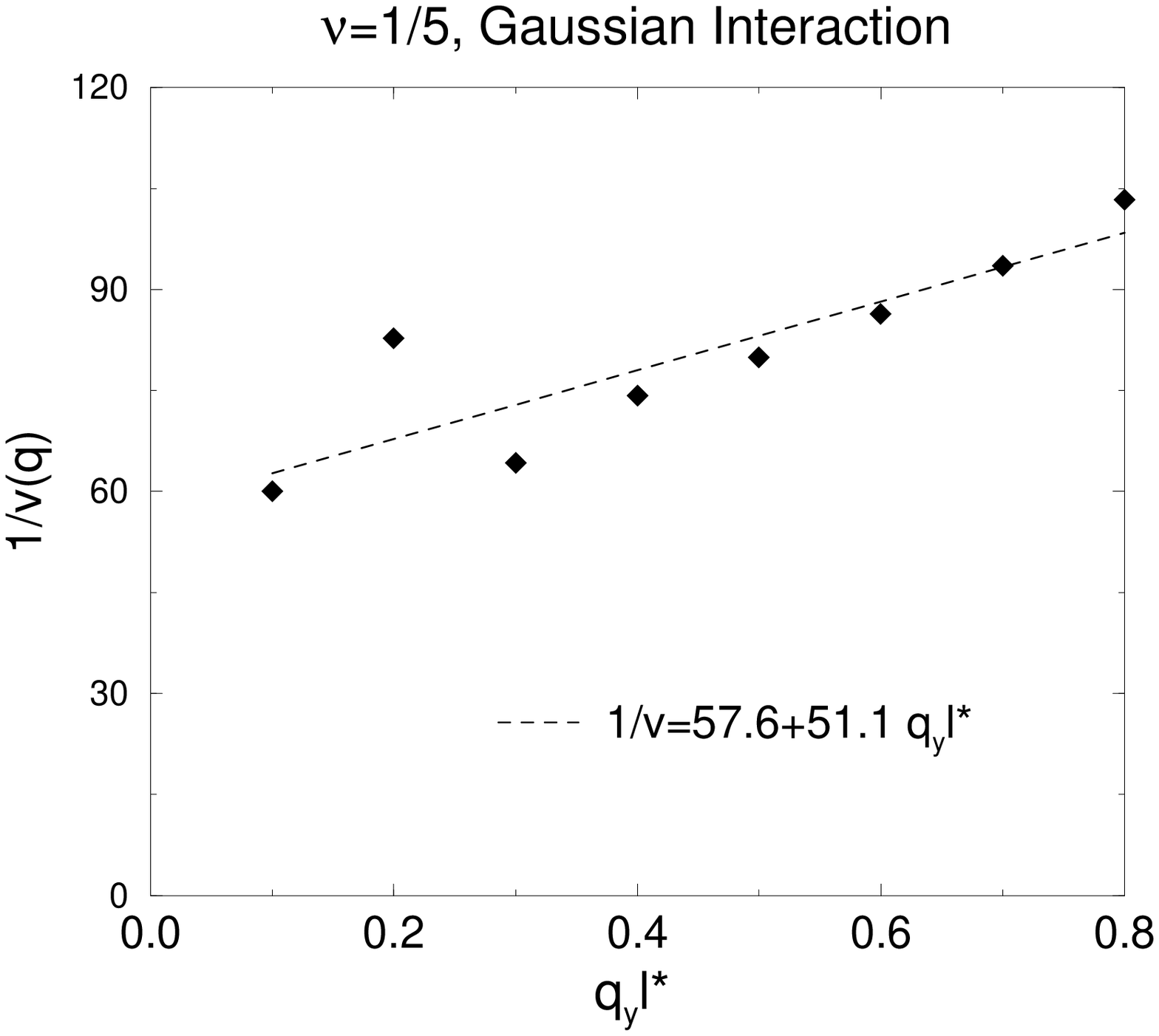}
\end{minipage}
\end{minipage}
\caption{Dispersion (left) and inverse CF velocity (right) for the $\nu=1/5$ 
edge collective mode with Gaussian interaction, $W=0$ and 10 CF-LLs. The fit 
of the inverse CF velocity (right) to our heuristic form is still reasonable. 
The large scatter at small momenta arises from uncertainties in locating the 
physical edge mode at small energies.}
\label{fig: dispersion.nu15}
\end{center}
\end{figure}

Next we consider filling factor $\nu=1/5$ which corresponds to CFs 
(consisting of electrons with $2s=4$ flux quanta) filling the lowest CF 
Landau level ($p=1$ and $c=\sqrt{4/5}$). Since there is no qualitative 
difference between collective edge modes obtained from Gaussian or 
Thomas-Fermi interactions, we concentrate on the Gaussian interaction which 
is more amenable to analytical calculations. 
Figure~\ref{fig: dispersion.nu15} shows the dispersion of the edge mode 
(left) and the inverse CF velocity (right), with $W=0$ and 10 CF-LLs. We note 
that the edge-mode energy scale is similar to that observed in the $\nu=1/3$ 
case. As expected, we find one edge mode and that mode softens with 
increasing $W$. We also find that the inverse CF velocity fits our heuristic 
form (\ref{eq: inverse_velocity}) and the free parameter $a$ has value 
similar to that in the $\nu=1/3$ case. 

\begin{figure}[b]
\includegraphics[scale=0.4]{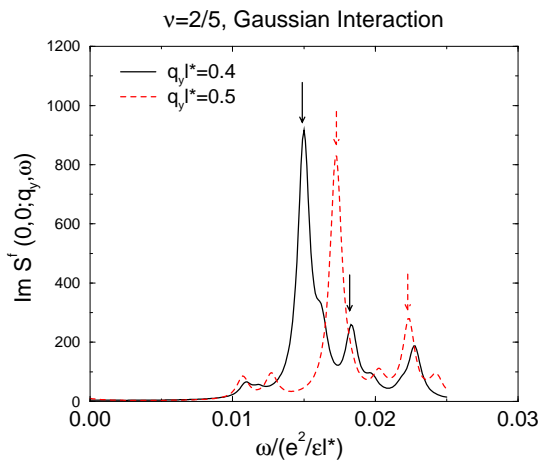}
\caption{Spectral function of the local electron density correlator for 
$\nu=2/5$ with Gaussian interaction, $W=0$ and 8 CF-LLs. The positions of 
the physical edge modes, defined by their stability with increasing CF-LLs 
and their positive energies, are marked with arrows for $q_yl^{*}=0.4$ 
(solid) and $q_yl^{*}=0.5$ (dotted).}
\label{fig: spectral.nu25}
\end{figure}
\begin{figure}[htb]
\begin{center}
\begin{minipage}{20cm}
\begin{minipage}{9cm}
\includegraphics[scale=0.4]{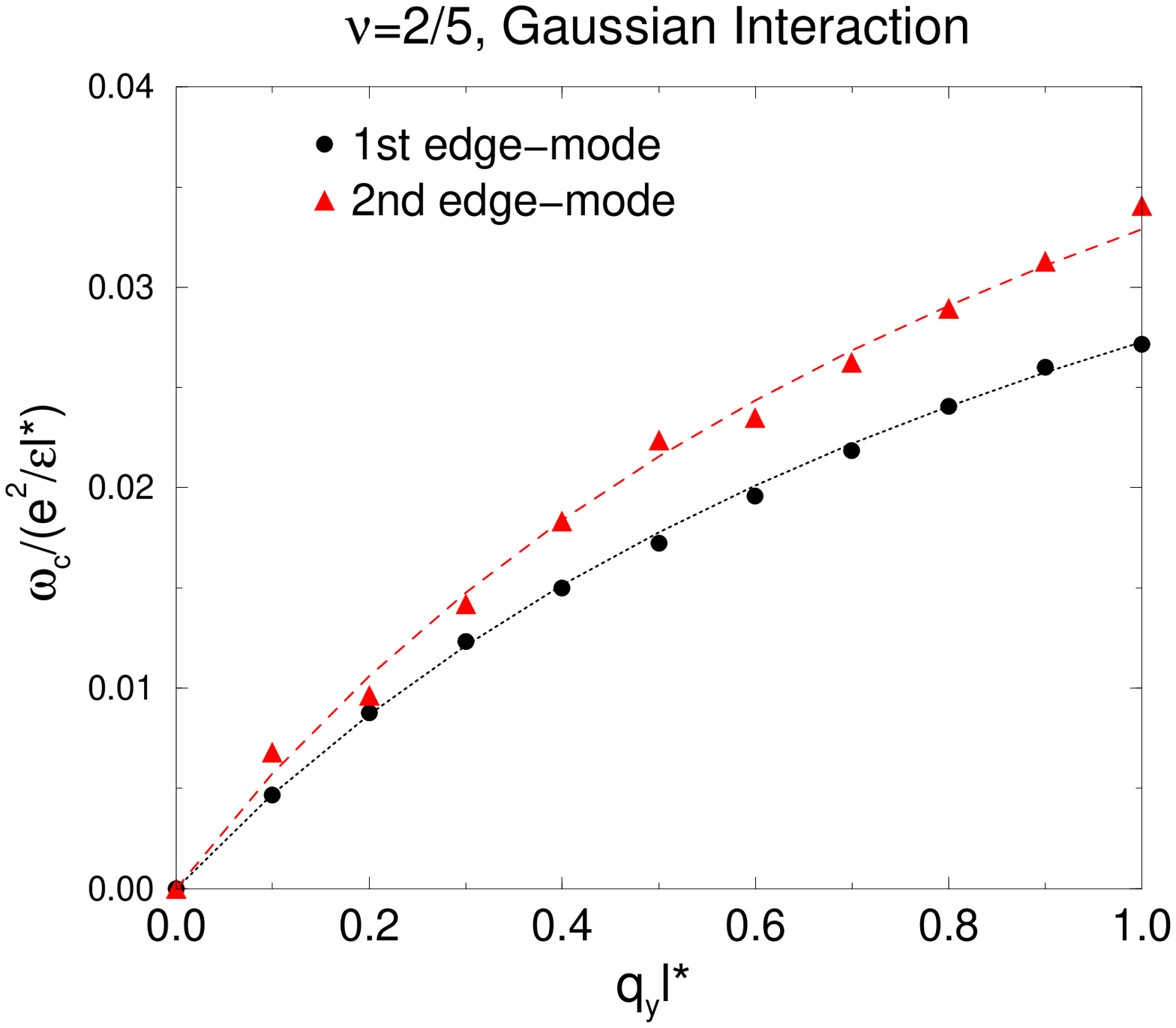}
\end{minipage}
\begin{minipage}{9cm}
\includegraphics[scale=0.4]{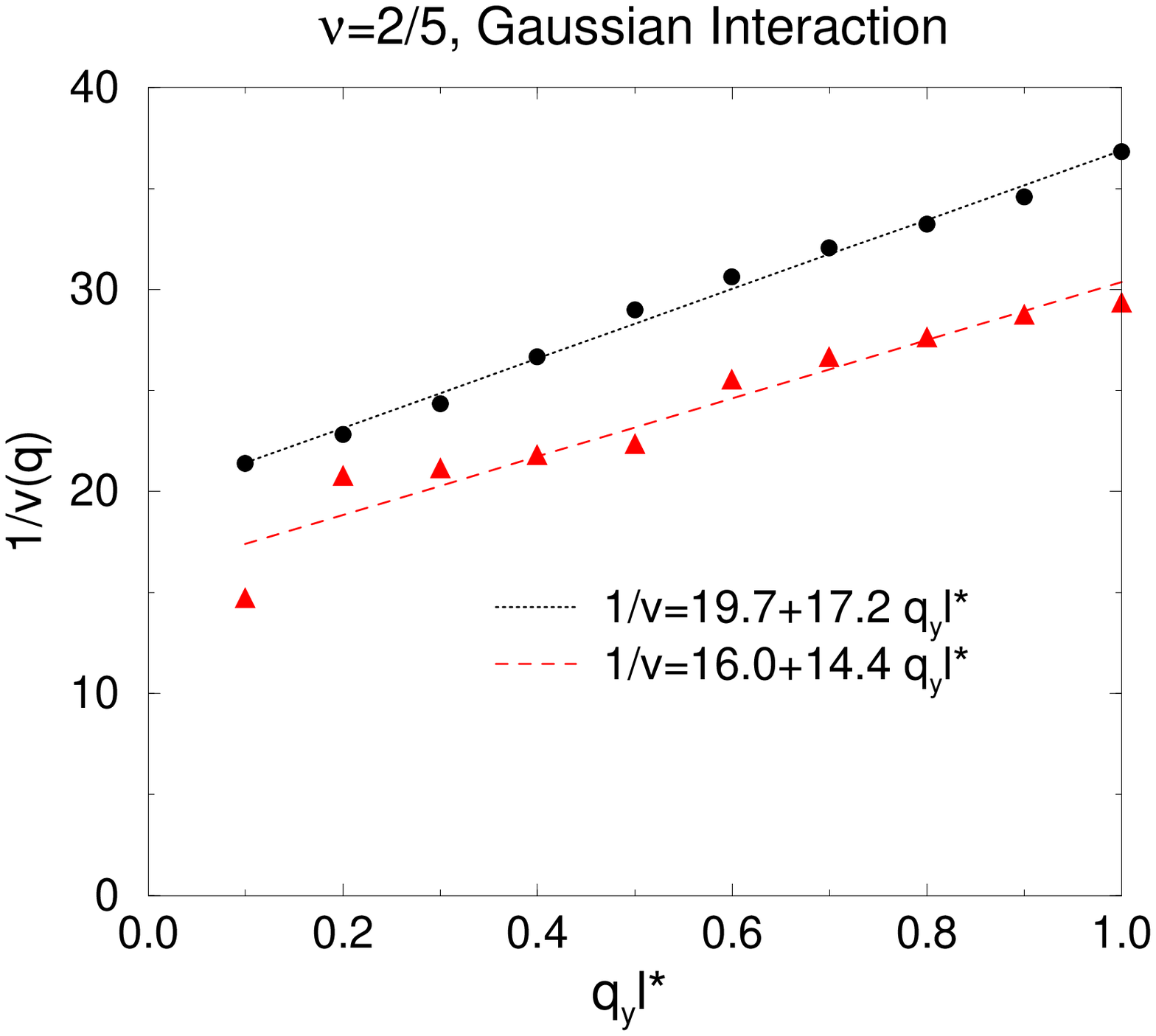}
\end{minipage}
\end{minipage}
\caption{Dispersions (left) and inverse CF velocities of edge modes for a 
$\nu=2/5$ system with Gaussian interaction, $W=0$ and 8 CF-LLs. The two 
modes have similar velocities and their inverse CF velocities fit the 
ansatz (\ref{eq: inverse_velocity}) reasonably well.}
\label{fig: dispersion.nu25}
\end{center}
\end{figure}

Let us now turn to the case of a fully spin-polarized $\nu=2/5$ state. Here 
we present, to the best of our knowledge, the first microscopic treatment of 
collective edge modes for $\nu=2/5$. In this case, the CFs (consisting of 
electrons with $2s=2$ flux quanta) fill $p=2$ CF Landau levels in the bulk, 
which leads to more involved numerical calculations. One now needs a TDHF 
matrix with twice the dimension as in the case of $\nu=1/3$ or $\nu=1/5$, 
since there are two filled CF-LLs in which the hole can be created. 
Consequently, we expect two gapless modes at the edge propagating in the same 
direction. Due to computational constraints, we restrict ourselves to 
Gaussian interaction and 8 CF-LLs. Figure~\ref{fig: spectral.nu25} shows the 
spectral function of the local density correlator, Im $S^{f}(0,0;q_y,\omega)$ 
for $q_yl^{*}=0.4$ and $q_yl^{*}=0.5$. For each value of $q_y$, two peaks 
marked by arrows represent the physical edge modes. It is interesting to note 
that contrary to the naive speculation, the edge mode with smaller energy has 
a greater spectral weight and therefore higher charge. Right panel in 
Fig.~\ref{fig: dispersion.nu25} shows dispersions for both edge modes, which 
have approximately the same velocity; the left panel shows that the inverse 
CF velocities can be fit to the heuristic form obtained earlier.

\section{Discussion}
\label{sec:conclusions}
\begin{figure}[htb]
\includegraphics[scale=0.4]{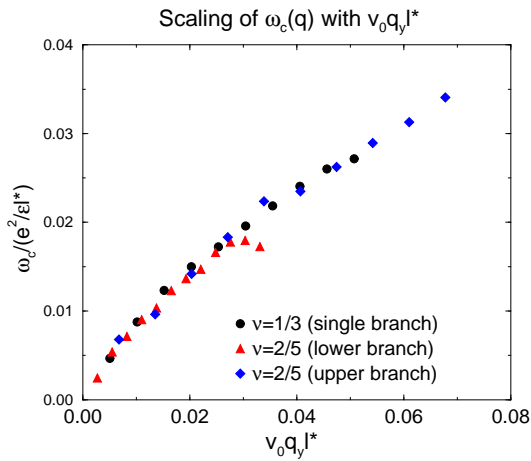}
\caption{Scaling of the single-branch of $\nu=1/3$ dispersion and both 
branches of the $\nu=2/5$ dispersion with $v_0$. Since both have $s=1$, and 
similar values of the free parameter $a$ which appears in 
Eq.(\ref{eq: inverse_velocity}), the data collapse on a single curve.}
\label{fig: scaling}
\end{figure}

In this paper we have presented a microscopic approach to calculating 
collective edge excitations in the {\it fractional} quantum Hall regime 
which is complementary to the exact diagonalization approach. This approach 
is based on the extended Hamiltonian theory~\cite{rmpus} and permits the use 
of many-body approximations which are generally valid in the {\it integer} 
quantum Hall regime. We have presented results for collective modes, and their 
dependence on the nature of electron-electron interaction and the width of 
the background confining potential at various filling factors. For an 
unreconstructed edge, at filling factor $\nu=1/3$ and $\nu=1/5$, we found a 
single linearly dispersing edge mode which softens with the increasing 
width $W$ of the confining potential. In contrast, we found two linearly 
dispersing modes for fully spin-polarized $\nu=2/5$ state. It is somewhat 
surprising that the two edge modes have nearly the same energy, and that the 
mode with the lower energy has a greater overlap with the charge operator. The 
curvature of the collective mode dispersion is somewhat unexpected from 
hydrodynamic theories. We show that it can be understood satisfactorily by 
assuming a momentum-dependent collective mode velocity, 
Eq.(\ref{eq: inverse_velocity}), which reflects the effect of decreasing 
electron density and the subsequent increase in the effective field $B^{*}$ 
seen by the CFs. Our ansatz implies that the magnetoexciton frequency 
$\omega_c(q)=v(q)q_y$ depends only on the combination $v_0q_y$. Incidentally, 
we find that the free parameter $a$ is roughly equal for $\nu=1/3$ dispersion 
and both branches of $\nu=2/5$ dispersion; therefore we can collapse all 
three dispersions (which have $s=1$) onto a single curve by scaling with 
$v_0$ (Fig.~\ref{fig: scaling}). Our results for edge-mode dispersions did 
not show the roton minimum near $q_yl^{*}\sim 1$ seen in exact 
diagonalization studies,\cite{wan} presumably because the system is not close 
to edge reconstruction.~\cite{yogesh}

Now let us mention some caveats. While our approach does not suffer from the 
computational limitations of exact diagonalization, we have computational 
limitations of our own. We work, in principle, in the thermodynamic limit. 
In practice, we are forced to truncate the filled states in the bulk and the 
empty states near the edge, to use a discretized Landau gauge index $X$, and 
to use a finite number of CF-LLs in the calculation. There are two 
limitations which we face: At the mean-field stage, we need to find the HF 
ground state as accurately as possible. Each step of the iterative HF 
calculation takes longer when these numbers get larger. In the TDHF stage, 
the main limitation is imposed by the time required to construct the TDHF 
matrix which involves a quadruple sum and increases as $N^4$ where $N$ is a 
measure of the number of single-particle levels kept. The diagonalization of 
the TDHF Hamiltonian takes a small fraction of the total computation time. We 
originally intended to study the softening of the collective edge-modes prior 
to edge reconstruction, and the evolution of edge-modes thereafter, as well. 
However, we were not successful in identifying the physical modes 
unambiguously at larger values of $W$. Presumably, with better computational 
resources, if we increase the number of CF-LLs used, this task will be 
feasible. 

What our approach does is to bring within computational reach the techniques 
which have been successful in the study of edge modes of {\it integer} 
quantum Hall systems.\cite{chamon,iqh-edge} It does have its limitations; 
despite these limitations, we believe it can be usefully employed for cases 
which cannot be studied microscopically by any other method. For example, if 
one wants to calculate the temperature dependence of edge mode dispersion, 
one requires the knowledge of all excited states in an exact diagonalization 
approach. This is computationally prohibitive compared to finding the ground 
state and a few excited states. The state of the art exact diagonalization 
studies can treat at most 7-9 particles if all the states are kept.\cite{wan} 
We expect the collective edge modes to be quite sensitive to temperature and 
finite size effects, since they are gapless. This may be the reason that 
although exact diagonalization studies with all states\cite{wan} reproduce 
the qualitative features of microwave absorption experiments,\cite{lloyd} the 
energy scales predicted are considerably higher than those in the 
experiments. Another important piece of physics, which our approach can 
easily handle, is the role of disorder. In any real sample there is bound to 
be disorder. If the dominant disorder is due to quenched fluctuations in the 
remote dopant layer, the system breaks up into incompressible strips 
separated by compressible regions.\cite{efros} These strips have been seen in 
imaging experiments.\cite{imaging} It is straightforward to include the 
disorder potential perturbatively in our formulation, both in the mean-field 
and the TDHF approximations. The HF matrix is somewhat more complicated due 
to possible spatial structure along the edge. However, using the clean 
single-particle and collective modes of the incompressible strip as a basis, 
one can construct a microscopic theory of collective edge modes with 
disorder. This will provide us with a microscopic approach for examining the 
parameters which enter hydrodynamic theories to this problem.\cite{kane} 

One can also address novel edge-states with our approach, such as an edge 
between $\nu=1/3$ and $\nu=2/5$ regions which have CFs with the same number of 
flux quanta or an edge between $\nu=1/3$ and $\nu=2/9$ regions, which have CFs
with different numbers of flux quanta attached. It would also be very 
interesting to study the edge dynamics of the Fermi-liquid-like state at 
$\nu=1/2$.\cite{eric}

\section*{Acknowledgments}
We thank Kun Yang for helpful discussions. This work was supported by the 
National Science Foundation under grant DMR-0311761 (HN and GM) and by the 
LDRD at Los Alamos National Laboratory (YJ).


\end{document}